\documentclass{PoS}
\pdfoutput=1

\usepackage{amsmath}
\usepackage{graphicx}
\graphicspath{{./Figs/}}

\usepackage[sort&compress,numbers,merge]{natbib}
\usepackage{natbibspacing}

\newcommand{\preprint}{
  \begin{picture}(0,0)
    \put(0,100){{\rm\normalsize HU-EP-13/18}}
  \end{picture}}

\title{\preprint%
Another look at the Landau-gauge gluon and ghost propagators at
 low momentum}

\ShortTitle{Another look at the Landau-gauge gluon and ghost propagators at
 low momentum}

\author{\speaker{Andr\'e Sternbeck}\thanks{Supported by the EU commission (IRG
256594).}\\
  Institut f\"ur Theoretische Physik, Universit\"at
  Regensburg, D-93040 Regensburg, Germany\\
  E-mail: \email{andre.sternbeck@ur.de}}

\author{Michael M{\"u}ller-Preussker\\
        Humboldt-Universit\"at zu Berlin, Institut f\"ur Physik,
  D-12489 Berlin, Germany}

\abstract{We study the gluon and ghost propagators of SU(2) lattice Landau
  gauge theory and find their low-momentum behavior being sensitive to the
  lowest non-trivial eigenvalue ($\lambda_1$) of the Faddeev-Popov operator. If
  the gauge-fixing favors Gribov copies with small (large) values for
  $\lambda_1$ both the ghost dressing function and the gluon propagator get
  enhanced (suppressed) at low momentum. For larger momenta no
  dependence on Gribov copies is seen. We compare our lattice data to the
  corresponding (decoupling) solutions from the DSE/FRGE study of Fischer, Maas
  and Pawlowski [Annals Phys.~324 (2009) 2408] and find qualitatively good
  agreement.}

\FullConference{Xth Quark Confinement and the Hadron Spectrum,\\
		October 8-12, 2012\\
		TUM Campus Garching, Munich, Germany}

\newcommand{\Fig}[1]{Fig.\,\ref{#1}}

\newcommand{\fc}{$f\!c$}
\newcommand{\lc}{$\ell c$}
\newcommand{\slc}{$s\ell c$}
\newcommand{\bc}{$bc$}
\newcommand{\hc}{$hc$}
\newcommand{\shc}{$shc$}

\begin{document}

\section{Introduction}

Studies of the elementary two and three-point functions of Landau-gauge
Yang-Mills theory have always been an interesting topic on the lattice. In
particular the gluon and ghost propagators became one in 2005 when it
was apparent that (continuum) functional methods
\cite{vonSmekal:1997is,*Zwanziger:2001kw,*Lerche:2002ep,*Zwanziger:2003cf} and
lattice approaches
\cite{Furui:2003jr,*Sternbeck:2005tk,*Boucaud:2005ce} disagree in their findings
for the propagator's low-momentum dependence. Since
then many efforts have been made, both on the lattice and in the
continuum, to verify and understand this discrepancy
\cite{Sternbeck:2007ug,*Cucchieri:2007md,*Cucchieri:2007rg,*Boucaud:2008ky,
*Boucaud:2008ji,*Sternbeck:2008mv,*Bornyakov:2009ug,*Bogolubsky:2009dc,
*Oliveira:2008uf,*Pawlowski:2009iv,*Oliveira:2012eh,*Fischer:2006ub,
*LlanesEstrada:2012my, Fischer:2008uz,Maas:2009se}.

A possible solution was proposed by Fischer, Maas and Pawlowski
in 2008 \cite{Fischer:2008uz}. They investigated the (truncated)
Dyson-Schwinger equations (DSEs) of the gluon and ghost propagators,
and in addition also the corresponding functional renormalization group
equations (FRGEs), and found there is not a unique solution to the system of
equations but a one-parameter family of \emph{decoupling} solutions. A
particular solution is chosen by the value set for the ghost dressing function
at zero momentum, $J(0)$. In the limit $J(0)\to\infty$, one obtains the
\emph{scaling} solution that was found before
\cite{vonSmekal:1997is,*Zwanziger:2001kw,*Lerche:2002ep}, but not yet on the
lattice.

Although this proposal may be attractive to understand the discrepancy, lattice
studies have not delivered evidence. In this contribution we show
that a part of this family of (decoupling) solutions
can be reproduced on the lattice, at least qualitatively, as far as possible on
a finite and rather coarse lattice, and to the extent computational resources
allow. We are also limited by our approach which allows only for mild variations
of the ghost dressing function at low
momenta. For this variation we utilize the lowest non-trivial eigenvalue
$\lambda_1>0$ of the Faddeev-Popov (FP) operator and show that the low-momentum
behavior of both the ghost dressing function and the gluon propagator changes
with the average $\lambda_1$ of the selected gauge-fixed (Gribov)
copies.\footnote{Alternatively, one could also directly constrain $J(p)$ at some
$p>0$ to select Gribov copies (see \cite{Maas:2009se}).}
Interestingly, these changes look qualitatively the same as for the
corresponding subset of decoupling solutions.

\section{Setup}

We restrict our study to SU(2) lattice gauge theory (Wilson plaquette action),
and also consider only one lattice size ($56^4$) and gauge coupling
($\beta=2.3$). This is fully sufficient for our purposes and allows us to
analyze (with reasonable computing time) data for the gluon propagator and the
ghost dressing function where their momentum dependence starts to
become flat, and this for a large enough number of Gribov copies such that a
correlation between $\lambda_1$ and the ghost and gluon propagators can be seen.
Moreover, by restricting to one lattice spacing and volume no further effects
(finite volume, renormalization, discretization) interfere. 

Our results are for an ensemble of 80 thermalized gauge field configurations.
These are separated by 2000 thermalization steps, each involving
four over-relaxation and one heatbath step. This number turns out to be
sufficient as no apparent autocorrelation effects are seen in the
data\footnote{At the conference results were presented for 60
configurations. To improve data and also to verify that our results do
not suffer from large autocorrelation effects, another (independent) chain of 20
configurations was generated. The results on either chain are fully compatible
with each other. A binning analysis is applied to
estimate statistical errors nonetheless.}. For every gauge configuration there
are at least $N_{\mathrm{copy}}=210$ gauge-fixed (Gribov) copies, generated
using an optimally-tuned over-relaxation algorithm. To ensure that these copies
are all distinct, the gauge-fixing always started from a random point on the
gauge orbit. Only a few Gribov copies were found twice. 

For each Gribov copy we calculate the lowest three (non-trivial) eigenvalues
$0<\lambda_1<\lambda_2<\lambda_3$ of the FP operator and use $\lambda_1$ to
classify copies. For each gauge configuration, the Gribov
copy with the lowest value for $\lambda_1$ is labeled \emph{lowest} copy
(\lc), while that with the highest $\lambda_1$ is called \emph{highest}
copy (\hc). The first generated copy, irrespective of
$\lambda_1$, gets the label \emph{first copy} (\fc). It represents an
arbitrary (random) Gribov copy of a configuration. To compare with
former lattice studies on the problem of Gribov copies we also
consider \emph{best copies}, i.e., those copies with the best (largest)
gauge functional value
\begin{equation}
 F_U[g] = \frac{1}{4V}\sum_{x}\sum_{\mu=1}^4\,
\mathfrak{Re}\operatorname{Tr}g_x U_{x\mu} g^\dagger_{x+\hat{\mu}}
\label{eq:Func}
\end{equation} 
for a particular gauge configuration $U\equiv\{U_{x\hat{\mu}}\}$. Here
$g\equiv\{g_{x}\}$ denotes one of the many gauge transformation fields fixing
$U$ to Landau gauge.

The gluon ($D$) and ghost propagators ($G$) are analyzed separately on
those four sets of Gribov copies. We apply standard recipes for
their calculation. For the ghost propagator, though, we will primarily analyze
its dressing function $J=p^2G$. It parametrizes the deviations
from the
tree-level (infrared diverging) propagator and is thus better suited for our
purposes. When quoting momenta in physical units we adopt the usual definition
$ap_{\mu}(k_{\mu}) = 2 \sin\left(\pi k_{\mu}/L_{\mu}\right)$ [with
$k_{\mu}\in(-L_{\mu}/2, L_{\mu}/2]$ and $L_{\mu}\equiv56$], assume for the
string tension $\sqrt{\sigma}=440\,\textrm{MeV}$ and use $\sigma a^2 = 0.145$
for $\beta=2.3$ from Ref.~\cite{Langfeld:2007zw}, where $a$ denotes the lattice
spacing. 

\section{Results}

\begin{figure*}[!t]
 \centering
 \mbox{\includegraphics[height=6.5cm]{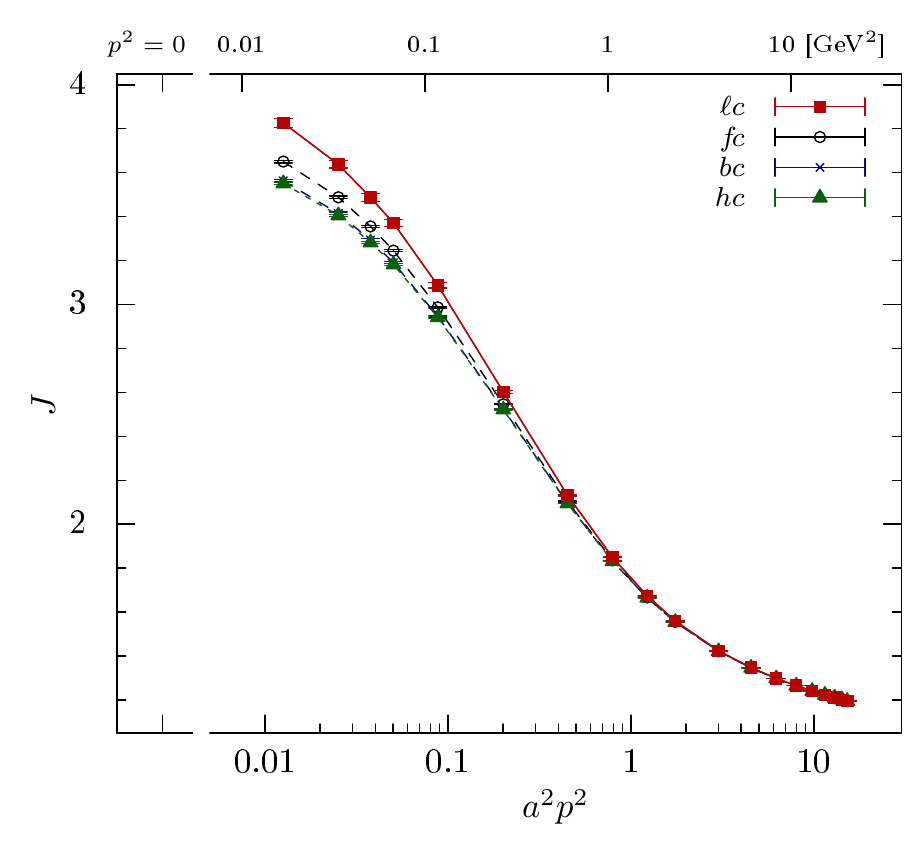}
 \includegraphics[height=6.5cm]{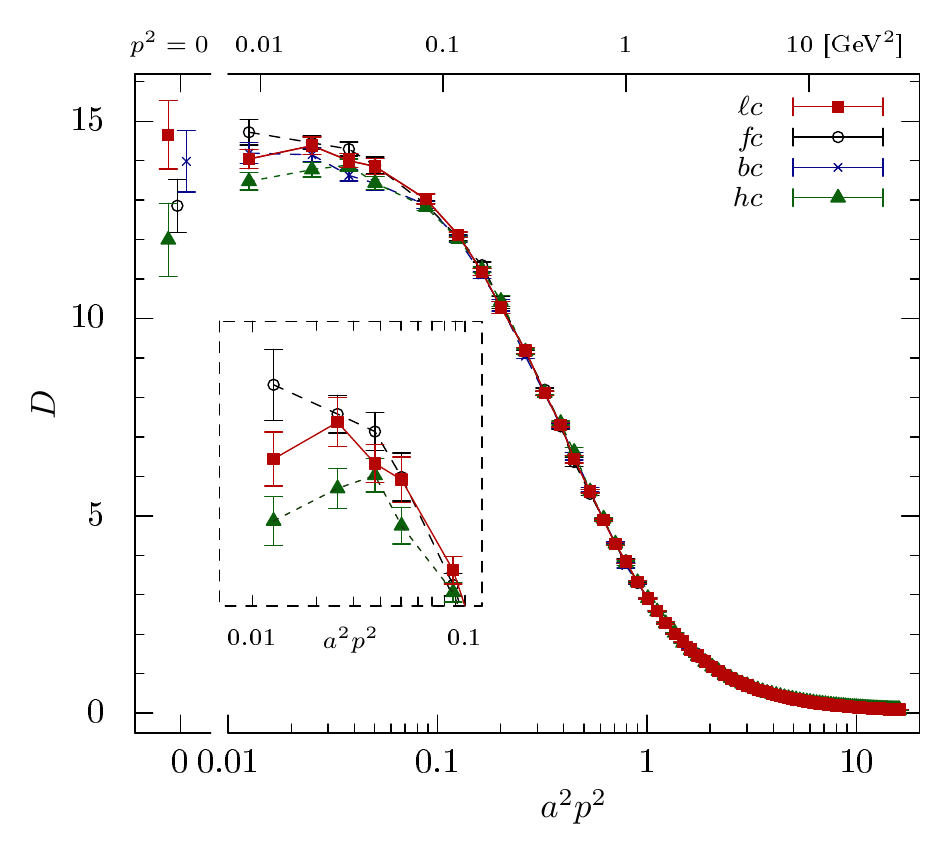}}
 \caption{Ghost dressing function (left) and gluon propagator (right)
    versus $a^2p^2$. Full squares (open circles, crosses, full triangles)
    refer to \lc\ (\fc, \bc, \hc) data; shown is
   the raw lattice data, that is no renormalization has been applied. A
   zoomed-in plot for the gluon propagator improves the visibility
   of the low-momentum region. For the same reason, points at $p=0$ are
   slightly shifted. Physical momenta are given at the top.}
\label{fig:ghdress_gl} 
\end{figure*}

Data for the ghost dressing function and the gluon propagator is shown
in \Fig{fig:ghdress_gl}. Looking there first at the left panel, one clearly
sees the choice of Gribov copies affects the momentum dependence of
the ghost dressing function at momenta $p^2\le0.2\,\text{GeV}^2$. Points for the
different sets of Gribov copies deviate systematically and the effect increases
the lower the momentum. For the \lc\ data we find the strongest enhancement
for the ghost dressing function towards zero momentum. The \hc\ data shows the
weakest enhancement, and this data also almost coincides with the \bc\ data (see
also the discussion below). On the other hand, for momenta above 1\,GeV no
Gribov-copy effects are seen.

Interestingly, also for the gluon propagator we see Gribov-copy effects
at the lowest momenta (right panel of \Fig{fig:ghdress_gl}). Due
to the larger statistical uncertainties (as typical for this propagator),
these effects are less significant however. A much enhanced statistics would
be desirable, but unfortunately this is beyond our
current resources of computing time. Nonetheless, a systematic deviation of the
\lc\ (red) and \hc\ data (green) is seen for momenta
$p^2<0.1\,\text{GeV}^2$, while for larger momenta the data points for all sets
agree within errors. We also see that the \bc\ data is suppressed compared to
the \fc\ data, in agreement with earlier studies
\cite{Bakeev:2003rr,*Bogolubsky:2007bw,*Bogolubsky:2009qb,*Bornyakov:2009ug}.

\begin{figure*}\centering
 \includegraphics[width=0.8\linewidth]{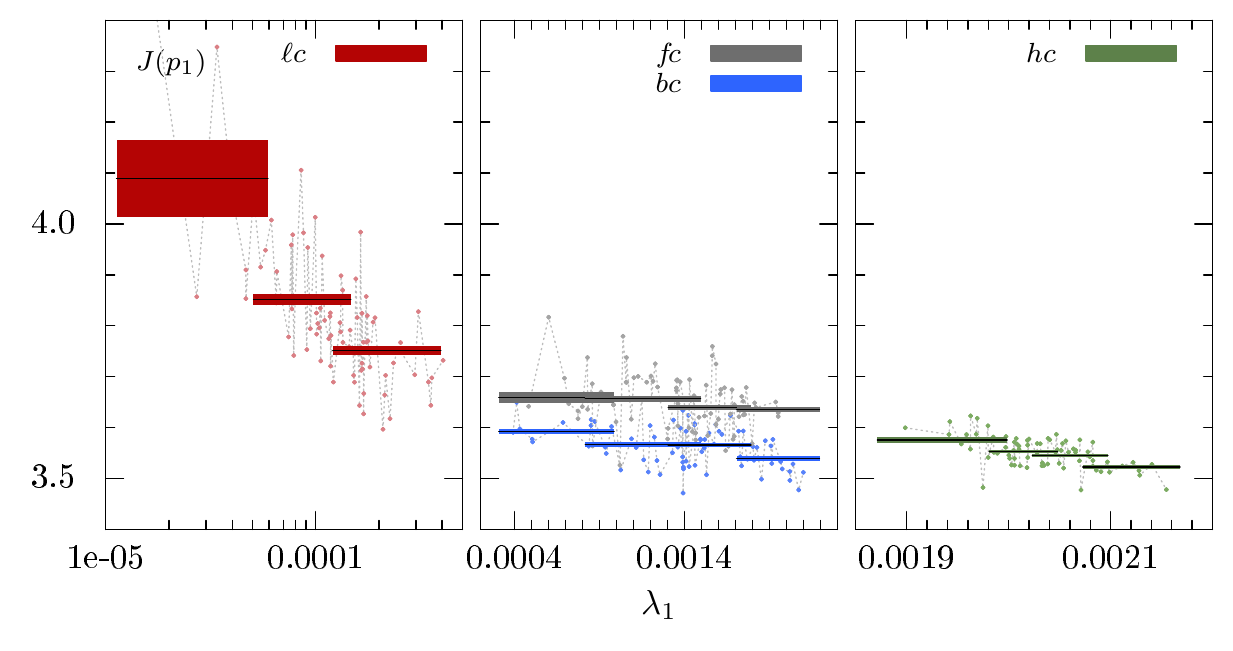}
 \includegraphics[width=0.8\linewidth]{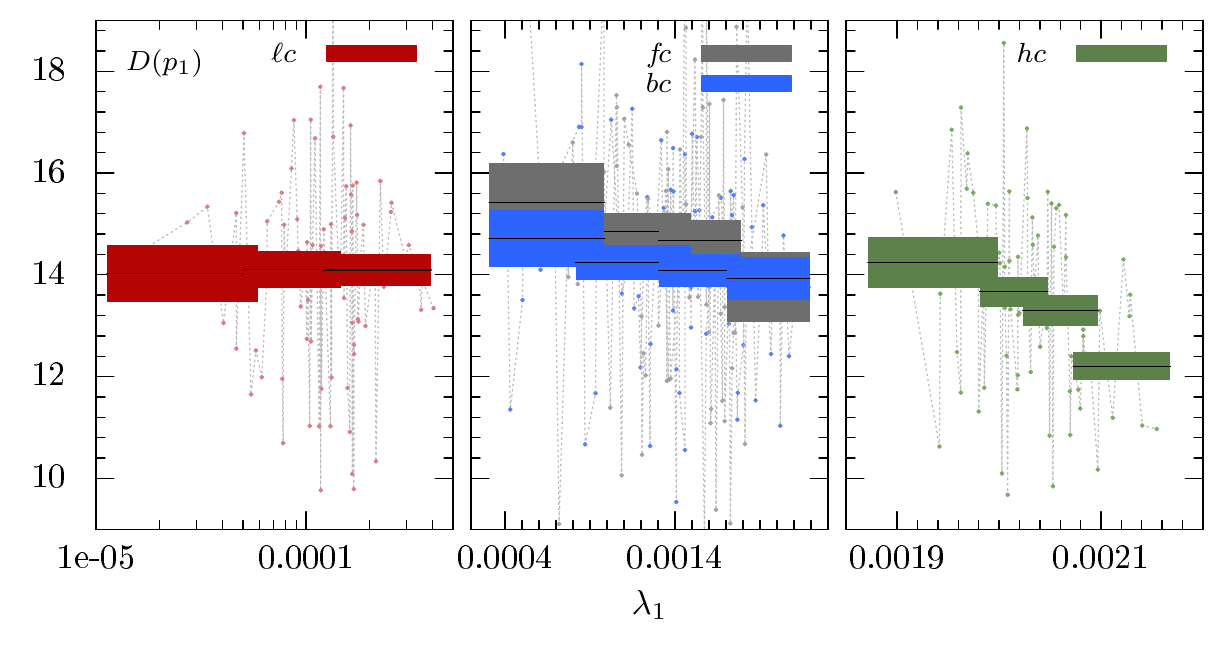}
 \includegraphics[width=0.8\linewidth]{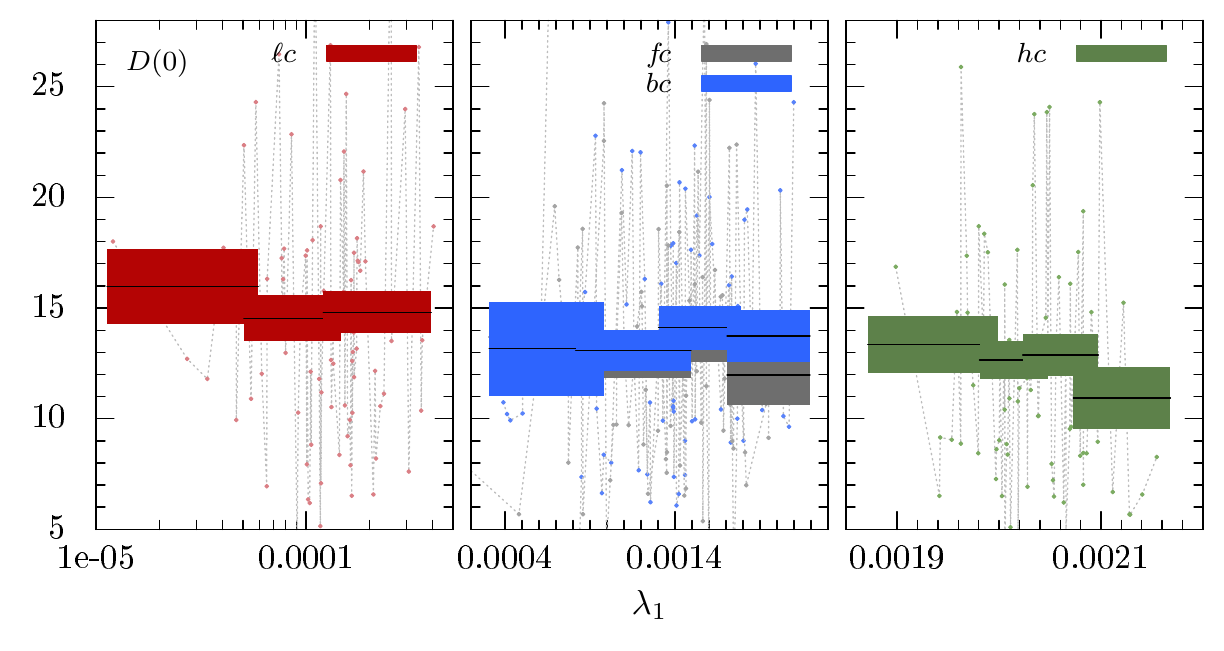}
\caption{Ghost dressing function (top) and gluon propagator at the lowest
finite momentum $p_1$ (middle) and zero momentum (bottom) versus $\lambda_1$. In
the background we show the ``measured`` values, separately for each Gribov copy
(scatter plot), and in the foreground  averages (colored bars). The middle line
of each bar marks the average over the shown $\lambda_1$-interval (bar
  length); the height reflects the statistical uncertainty. Note the different
  scales of the ordinates, and the partly
  overlapping ranges of $\lambda_1$ for the three panels showing data for our
 different sets Gribov copies (first, lowest, best and highest copies:
 \fc, \lc, \bc\ and \hc).}
 \label{fig:gh_p1_fps_gl_p1_fps}
\end{figure*}

The correlation between the propagator's low-momentum behavior and $\lambda_1$
is even better seen when looking at the values for $\lambda_1$ and the
corresponding (``measured'') values for the propagators, that is separately for
each Gribov copy. Such scatter plots are shown in \Fig{fig:gh_p1_fps_gl_p1_fps};
the
top panel shows the ghost dressing function at the lowest finite momentum
($a^2p^2_1\approx 0.01258$) and the other two the gluon propagator at $a^2p^2_1$
(middle) and at $p^2=0$ (bottom). To increase visibility of the $\lambda_1$
dependence, we also show averaged values (colored bars) over adequately chosen,
partly overlapping $\lambda_1$ intervals. The bar length equals the $\lambda_1$
interval and the bar height reflects the statistical uncertainty of each average
(marked by a line). 

For the ghost dressing function we clearly see the data points to grow if
$\lambda_1$ is decreased and vice versa. In particular towards small $\lambda_1$
the effect becomes large. We also find (middle panel) the \fc\ points
to lie above the \bc\ points, as expected from other studies, but this
dependence on the gauge-functional value adds to the dependence on $\lambda_1$.
This may explain why the \bc\ and \hc\ ghost dressing functions
(accidentally) coincide as noticed above. 

In comparison, the gluon propagator data comes with much larger statistical
fluctuations which makes it hard to draw finite conclusions.
Nonetheless, a trend is seen in the data: for large $\lambda_1$ the data points
tend to lower values than for smaller $\lambda_1$. Though, it is hard to decide
if this trend persists for very small $\lambda_1$. From the middle
and bottom panels of \Fig{fig:gh_p1_fps_gl_p1_fps} one could conclude either.
Note that this ambivalence is also reflected in the gluon propagator data
shown in \Fig{fig:ghdress_gl}. There we see the order of the \lc\ and \fc\
points at $p=0$ and low $p>0$ being partly inverted. Though, given the
findings below (\Fig{fig:gl_qq_lchc_2lc2hc}), this may be just a
statistical artifact. 

Let us remind that our averages for $D(0)$ result from about 80 individual
''measurements``, one per gauge copy, while each Monte-Carlo history point for
$D(p>0)$ itself is an average over all possible momentum directions with same
$a^2p^2$. For the lowest finite $p^2$ there are already 4 directions one
averages over. This explains the larger statistical noise for $D(0)$.

\begin{floatingfigure}[r]
 \centering
 \parbox{8.cm}{%
    \includegraphics[width=1.04\linewidth]{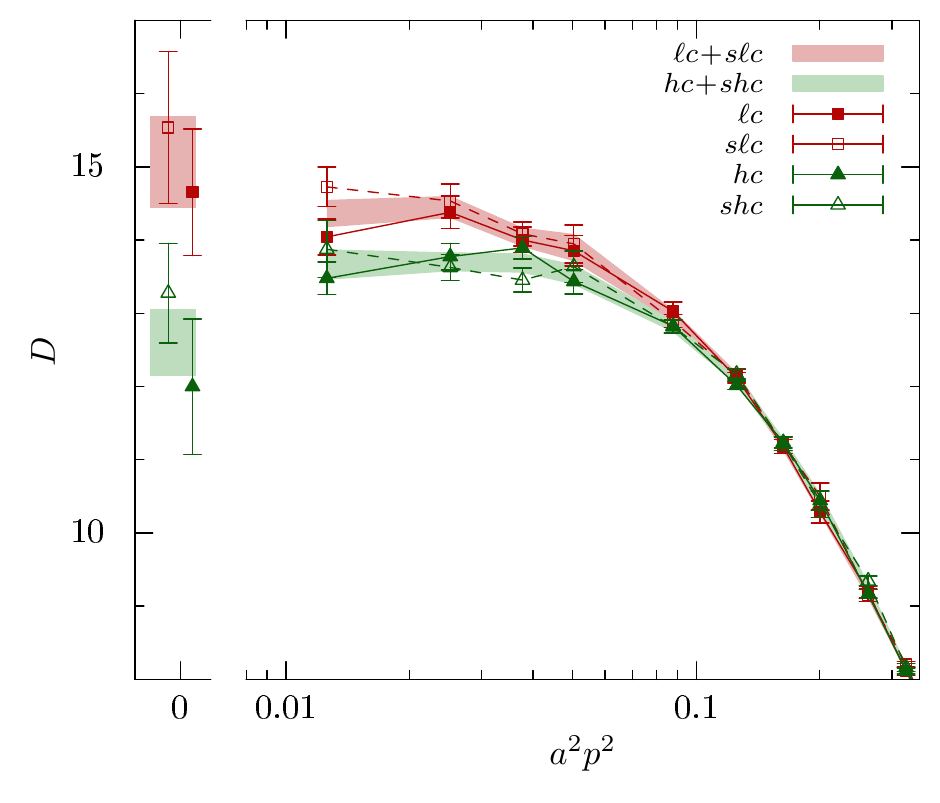}
    \caption{Gluon propagator versus $a^2p^2$. Data is shown separately
    for the sets of Gribov copies with lowest (\lc), second lowest (\slc),
    highest (\hc) and second highest (\shc) value for $\lambda_1$. Error bands
    in the background are for the combined \lc\ and \slc\ data (red) and
    the combined \hc\ and \shc\ data (green). At $p=0$ points are
    slightly shifted.}}
\label{fig:gl_qq_lchc_2lc2hc} 
\end{floatingfigure}
To cross-check if the $\lambda_1$-dependence we see for the gluon
propagator is not just a statistical artifact, we perform additional
calculations of the gluon propagator on all Gribov copies with
the second lowest and second highest $\lambda_1$. Those additional sets of
($2\times 80$) Gribov copies (labeled \slc\ and \shc\ in what follows) are
distinct from the sets of lowest and highest copies (\lc\ and \hc) analyzed
above, and if there is a dependence on $\lambda_1$, one should also see it when
comparing \slc\ and \shc\ data. And in fact, also this data clearly
exhibits a $\lambda_1$-dependence at low momenta (see
\Fig{fig:gl_qq_lchc_2lc2hc}). The combined \lc\
and \slc\ data and the combined \hc\ and \shc\ data (colored error bands in
\Fig{fig:gl_qq_lchc_2lc2hc}) currently give the
best impression of this dependence. Note that such a combination of data is
justified, as there are no correlations visible between data from different
copies of the same configuration, and by construction these sets of Gribov
copies come also with similar small or large values for $\lambda_1$: The
averaged $\lambda_1$ values on these four sets of Gribov copies are: 
$\langle\lambda_1\rangle_{\ell c}=1.43(9)\times10^{-4}$,
$\langle\lambda_1\rangle_{s\ell c}=2.14(9)\times10^{-4}$,
$\langle\lambda_1\rangle_{hc}=20.33(6)\times10^{-4}$ and
$\langle\lambda_1\rangle_{shc}=19.89(4)\times10^{-4}$ (lattice units).

\section{Conclusion}

\begin{figure*}
 \centering
 \includegraphics[width=0.9\linewidth]{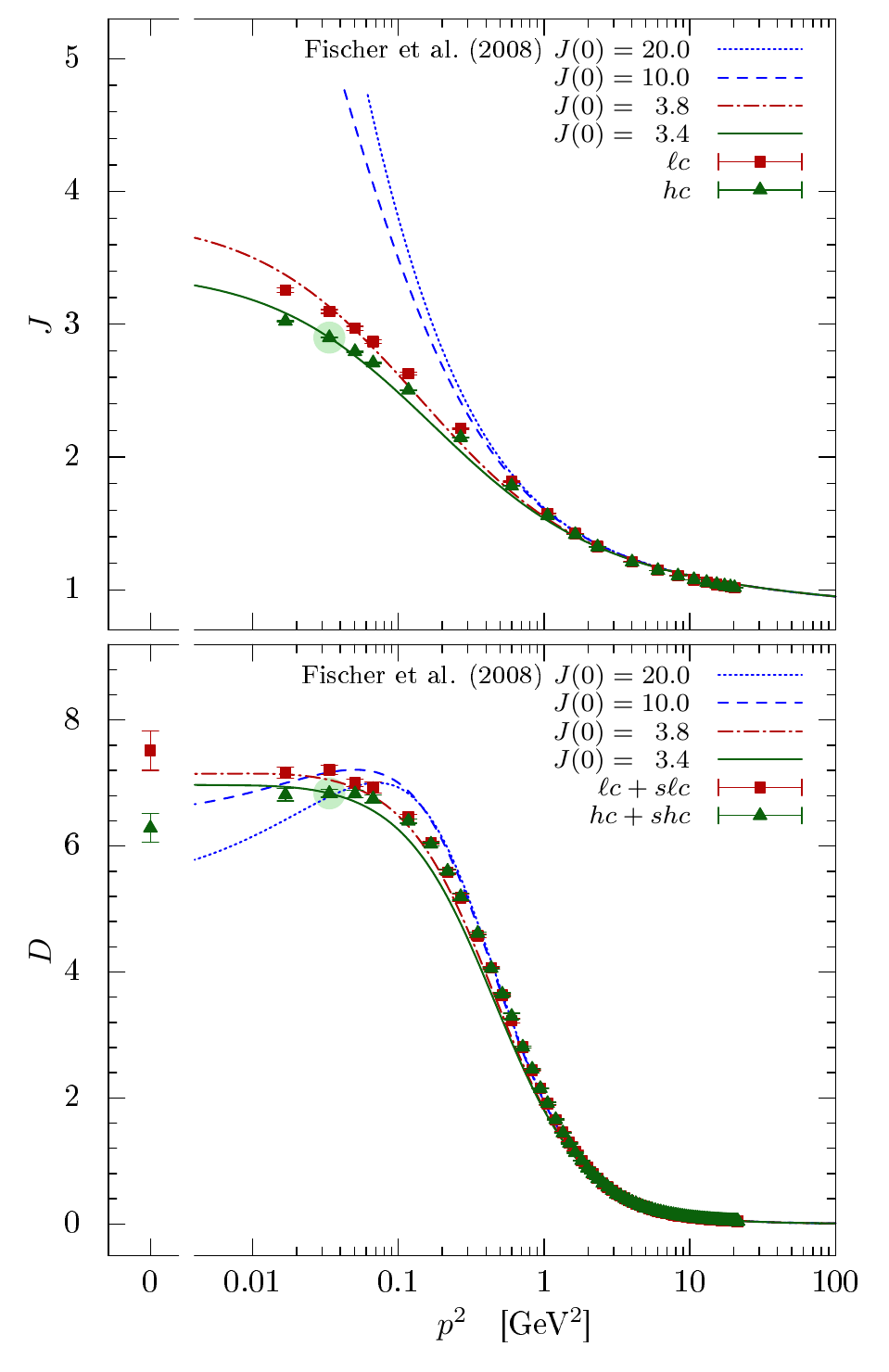}
 \caption{Ghost dressing function (top) and gluon propagator (bottom)
    versus $p^2$. Full symbols refer to our lattice data and lines to
    four selected decoupling (DSE) solutions from \cite{Fischer:2008uz}. Note
    that the order of the gluon propagator lines at low momenta changes
    somewhere between $J(0)=3.8$ and 10.}
\label{fig:gl_ghdress_qq_lc2lc_hc2hc_DSE} 
\end{figure*}

Our lattice study shows that the low-momentum behavior of the Landau-gauge gluon
and ghost propagators can be changed (slightly, though systematically) on the
lattice by constraining the lowest non-trivial eigenvalue
$\lambda_1$ of the FP operator. If we restrict $\lambda_1$ to be small (large)
for each Gribov copy, the ghost propagator at low momentum gets enhanced
(suppressed), while it is not at all affected at intermediate or large
momentum. Interestingly, also the gluon propagator ($D$) is affected at low
momentum, but in comparison to the ghost dressing function the effect is
smaller. Currently the effect is best seen if one combines the data from Gribov
copies with lowest and second lowest $\lambda_1$, and from copies with highest
and second highest $\lambda_1$ as shown, e.g., in
Figs.~\ref{fig:gl_qq_lchc_2lc2hc} or \ref{fig:gl_ghdress_qq_lc2lc_hc2hc_DSE}. 

In \Fig{fig:gl_ghdress_qq_lc2lc_hc2hc_DSE} we also confront
our data for the ghost dressing function and the gluon propagator with a
corresponding pair of decoupling solutions from \cite{Fischer:2008uz}. These
DSE solutions are approximately those where the boundary condition on the ghost
dressing function was set to $J(0)=3.4$ and $J(0)=3.8$, respectively. For the
comparison our data has been renormalized relatively to the given decoupling
solution, by applying a common renormalization factor ($Z_J$ and $Z_D$) to the
respective data. Since truncation effects are expected to become important,
these two factors were chosen such that the \hc-data points (green triangles)
agree with the $J(0)=3.4$ curves (green) at the second lowest finite momenta
(this point is highlighted by green circle in the figure). One could of
course chose any other point, which would result in a similar figure. But
at the moment our comparison is only qualitative anyway. Nonetheless, the
surprisingly good agreement between the so different approaches to the gluon and
ghost propagators of Landau-gauge Yang-Mills theory is encouraging. 

More details will became available soon in a revised version of
\cite{Sternbeck:2012mf}.

\section*{Acknowledgments}

This work was supported by the European Union under the Grant Agreement number
IRG 256594. We thank C.~Fischer, A.~Maas and J.~Pawlowski for discussions and
for providing us (partly unpublished) information and data. We acknowledge
generous support of computing time by the HLRN (Germany).


%

\end{document}